# Sequential Operating Simulation of Solid State Transformer-Driven Next-Generation 800 VDC Data Center

Jian Xu, *Senior Member, IEEE*, Xinxiong Jiang*, *Member, IEEE*, Yi Bao, Yuchen Zheng, Xuhui Chen, Qiang Xu, Siyang Liao, *Member, IEEE*, Deping Ke, and Xiaoqi Gao

*Abstract*—Artificial-intelligence (AI) workloads are driving rapid growth in data-center electricity use and rack power density, increasing demand for power-delivery systems that are efficient and robust to fast load transients. Conventional uninterruptible power supply (UPS) based AC distribution chains involve multiple conversion stages and line-frequency transformers, which compound losses and are less compatible with dynamic AI power profiles. Although solid-state transformers (SSTs) and 800 VDC distribution architecture are widely discussed, implementable topology/control details, and long-horizon validation with realistic operating profiles remain limited. This paper develops an SST-driven 800 VDC architecture that converts 10 kV MVAC to an 800V LVDC bus using a three-phase H-bridge AC/DC stage cascaded with a dual-active-bridge (DAB) DC/DC stage. A coordinated closed-loop control scheme, combining rectifier voltage/current regulation and DAB phase-shift control, is designed to maintain DC-bus voltage stability. The proposed system is implemented on the real-time digital simulation (RTDS) platform and evaluated via sequential simulations using real-world day- and month-scale operating profiles of data centers, benchmarked against a UPS supply chain. Numerical studies demonstrate tight 800 VDC regulation, reduced input-side energy consumption compared with the UPS baseline, and satisfactory power-quality performance. A capacitance sensitivity test quantifies tradeoffs between DC-bus ripple and low-frequency input-power oscillations, yielding a practical capacitance range for design. Overall, the work provides a reproducible evaluation workflow and actionable guidance for next-generation AI data centers.

## I. INTRODUCTION

With the rapid expansion of AI, global computational demand has surged, driving an unprecedented rise in data-center electricity consumption. Recent assessments show that AI-accelerated servers have become a dominant contributor to this growth, causing U.S. data-center electricity use to more than double from 2017 to 2023, reaching 176 TWh and representing 4.4% of total national electricity consumption [1, 2]. Looking forward, projections indicate that data-center demand could rise to 325-580 TWh by 2028 in the U.S. alone,

Jian Xu, Xinxiong Jiang (*Corresponding Author*), Yuchen Zheng, Xuhui Chen, Qiang Xu, Siyang Liao, and Deping Ke are with the School of Electrical Engineering and Automation, Wuhan University, Wuhan 430072, China (e-mail: xujian@whu.edu.cn; jiangxinxiong@whu.edu.cn; zhengyuchen0601 @whu.edu.cn; chenchenchen@whu.edu.cn; qiangxu@whu.edu.cn; liaosiyang @whu.edu.cn; kedeping@whu.edu.cn).

Yi Bao and Xiaoqi Gao are with the Energy Innovation Department, Beijing VNET Broadband Data Center Co., Ltd, Beijing 100006, China (e-mail: bao.yi@vnet.com; gao.xiaoqi@vnet.com).

reflecting both rapid AI hardware deployment and intensifying cooling requirements [3]. Globally, similar trends are emerging, with multiple analyses identifying data centers and AI as major new drivers of electricity demand in the coming decade [4]. More than 85% of new capacity is expected to cluster in the United States, China, and the European Union, placing additional strain on already congested grids [5]. As a result, the shift toward higher power-density IT equipment and increasingly energy-intensive AI workloads is creating an urgent need for more efficient, high-capacity power distribution architectures within modern data centers.

Conventional data centers are predominantly powered through AC distribution architectures centered around UPS systems, where utility power is stepped down and conditioned by double-conversion UPS units before being delivered to IT racks [6]. While this architecture has been the industry standard for decades, it presents several structural limitations that hinder its ability to support modern AI-oriented computing environments. Traditional UPS systems rely on large line-frequency transformers and multiple conversion stages, resulting in significant cumulative losses and leaving limited room for further efficiency improvement. In addition, double-conversion topologies operate continuously through rectifier and inverter stages, which further depresses efficiency at part load [7]. The AC-based power path is inherently unidirectional and optimized for relatively steady loads, making it poorly suited for the fast transient behavior and steep power fluctuations characteristic of AI accelerator workloads. These systems also exhibit relatively slow dynamic response and cannot effectively track rapid load changes, leading to oversizing of UPS modules and distribution equipment to maintain reliability margins [8]. As AI clusters push rack power densities to unprecedented levels, the inefficiency, limited responsiveness, and architectural rigidity of conventional UPS-based AC distribution increasingly become fundamental barriers to scalable and energy-efficient data-center power delivery [9].

To address the limitations of the traditional AC UPS system, the data-center industry has begun transitioning toward new power-delivery paradigms such as high-voltage direct current (HVDC) power supply architectures and Panama-style hybrid AC/DC power supply architectures. These emerging schemes reduce the number of conversion stages, enhance power-delivery efficiency, and better accommodate the high and rapidly varying power demands of AI workloads. HVDC power supply architectures improve distribution efficiency by using high-voltage DC transmission inside the facility [10], while Panama power supply architectures integrate both AC and DC distribution paths to achieve a balance between compatibility with existing infrastructure and the efficiency benefits of DC distribution [11-13].

Building upon these innovations, the most prominent development trend is the adoption of SST technology, which is increasingly recognized by major data-center and power-equipment vendors as a next-generation cornerstone for AI-driven power architectures [14, 15]. Over the past decades, medium-voltage (MV) SSTs have also seen rapid global development in a wide range of applications, including high-speed railway traction systems, extreme fast-charging infrastructure, grid-tied renewable-energy generation, and urban smart grids, thereby demonstrating their maturity and scalability across mission-critical environments [16, 17]. SSTs employ high-frequency isolation and fully electronic conversion stages to replace bulky line-frequency transformers, resulting in significantly reduced volume, enhanced controllability, and improved dynamic performance [18, 19]. Their capability to flexibly transform between different electrical forms within a unified platform enables seamless interfacing with medium-voltage (MV) feeders while providing regulated low-voltage (LV) outputs suitable for rack-level distribution [20]. This makes SSTs particularly advantageous for supporting high power density, bidirectional power flow, and fast response to rapid load fluctuations characteristic of AI computing clusters [21]. As data centers evolve toward higher power density and software-defined electrical infrastructures, SST-based architectures are rapidly emerging as the dominant direction for next-generation AI data-center power systems [22].

Although recent white papers and industry reports highlight the promising role of SSTs in next-generation data-center power architectures, they generally stop short of explaining how such SST systems are actually realized. Current publications do not specify which SST topologies are suitable for data-center deployment, nor do they present the detailed design parameters, control architectures, or component-level requirements needed for practical engineering implementation. Furthermore, the efficiency figures, harmonic characteristics, and power-quality metrics reported in these documents are typically not accompanied by transparent methodologies or validation procedures, leaving a gap in understanding how these values were obtained and how they compare to existing power-conversion architectures at the data-center level [6, 23].

Moreover, AI-driven data centers require evaluation across a much broader temporal spectrum, including long-duration operational performance, load-following capability, and real-world interaction with highly nonuniform AI computational profiles [24]. Existing studies on DC microgrids and DC distribution systems largely focus on steady-state performance or relatively short-term dynamics for buildings, commercial facilities, and generic mission-critical systems [8, 10, 13]. Only a limited number of works explicitly consider the combination of high-density AI loads, fast-ramping behavior, and month-scale energy-management implications in data-center environments [25, 26]. In parallel, the progressive shift toward DC-operated environmental-control subsystems in data centers—including cooling, pumping, and airflow-regulation equipment—is often discussed in DC-building and microgrid literature but is rarely integrated into SST-oriented studies [12, 27, 28]. As a result, there is a lack of co-simulation frameworks that jointly evaluate SST-based power delivery, DC environmental-control systems, and realistic AI workload dynamics, limiting our understanding of holistic system performance in future AI data-center environments [25].

In this work, a three-stage SST-driven power architecture tailored for next-generation 800 VDC data centers is developed, based on a three-phase H-bridge AC/DC rectifier cascaded with a DAB DC/DC converter. The proposed SST system directly converts 10 kV utility MVAC to 800V LVDC, substantially reducing intermediate conversion stages and improving overall energy-conversion efficiency. On this basis, a coordinated control framework is designed: the three-phase H-bridge rectifier employs an outer voltage-control loop and an inner current-control loop, combined with carrier-based PWM and gating logic to regulate the DC-link capacitor voltage. Meanwhile, the DAB converter utilizes a phase-shift modulation strategy to achieve closed-loop voltage regulation at the 800 VDC bus, ensuring adaptive and stable performance under rapid load variations.

To reflect the evolving characteristics of AI-oriented data centers, the 800 VDC bus is further integrated with DC environmental-control loads, battery energy storage, and distributed photovoltaic generation, forming a comprehensive compute–cooling–power co-simulation model. This model is implemented on a Real-Time Digital Simulator (RTDS) platform, where the fast-varying AI computing demand is represented through dynamic impedance equivalents. Moreover, a monthly continuous operating profile collected from a real data center in Taicang, Jiangsu, is incorporated to perform long-duration sequential simulations. Under different SST parameter settings and resource-integration scenarios, we evaluate monthly energy efficiency, harmonic performance, and voltage deviation, demonstrating the high efficiency, low distortion, and robust voltage-regulation capability of the proposed SST-driven architecture, and uncovering several insights regarding the behavior of future 800 VDC data-center systems. This paper describes the following contributions of this work:

● A complete SST model based on a three-phase H-bridge AC/DC rectifier and DAB DC/DC converter is constructed, and instantiate a full 800 VDC data-center architecture, including DC environmental-control systems, distributed PV, and battery storage, enabling holistic emulation of next-generation AI data-center energy structures.

● Using an RTDS high-fidelity real-time simulation platform, we integrate closed-loop control for both the three-phase H-bridge AC/DC stage and the DAB DC/DC stage, achieving stable 800 VDC regulation and enabling daily-scale and month-scale sequential simulation that reproduces realistic operating behaviors of an 800 VDC data center.

● Based on long-duration sequential simulation, several key findings are uncovered regarding system efficiency, power quality, and operational characteristics under diverse SST settings and resource-integration conditions. Compared with the conventional UPS architecture, the SST-based power supply provides more reliable DC-bus voltage and achieves significant energy savings across different load types and time scales. Moreover, within a certain range, increasing the low-voltage side capacitance of DAB can effectively suppress the DC-bus voltage ripple, as well as the energy loss and harmonics.

The rest of this paper is organized as follows: The developed SST-driven architecture for 800 VDC data centers is introduced in Section II. Section III presents its modeling implementation in RTDS, and the corresponding control patterns are designed. Simulation studies are performed in Section IV. Finally, conclusions are drawn in Section V.

## II. Solid State Transformer-Driven Architecture for 800 VDC Data Center

### A. Overall Architecture

The 800 VDC data center power supply architecture is shown in the Fig. 1. The system is built around an SST and features a distributed power-processing framework characterized by fully bidirectional and controllable energy flows. The utility AC is converted by a high-frequency, modular cascaded SST stage, where it undergoes rectification, galvanic isolation, and regulation to form a stable and controllable low-voltage DC bus. Subsequently, each SST module performs coordinated control according to load demand, enabling flexible power distribution, rapid voltage regulation, fault isolation, and parallel/auto-current-sharing operation. Ultimately, the downstream interfaces deliver high-quality, low-harmonic, and real-time-adjustable power to the loads. Compared to conventional line-frequency transformer architectures, this topology provides high power density, high efficiency, bidirectional energy transfer, and inherent compatibility with distributed energy resources and energy-storage systems throughout the entire conversion path from the utility grid to the load.

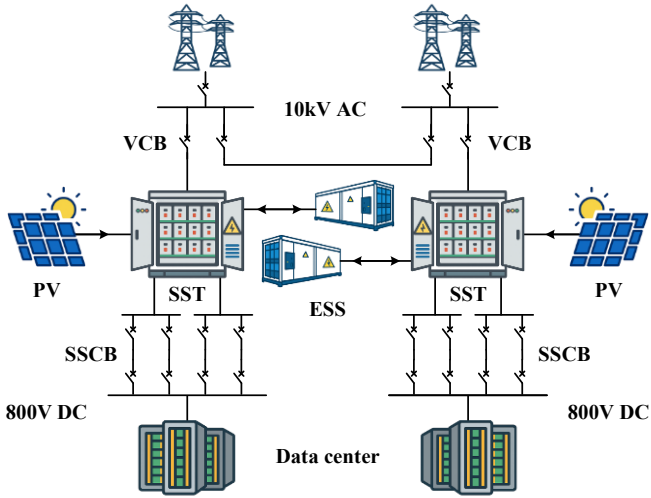

**Fig. 1.** The overall topology of SST-driven 800 VDC data center.

### B. Solid State Transformer (SST)

The main circuit schematic of the SST is shown in Fig. 2. This circuit consists of the AC-DC converter and the DAB converter. The DAB converter is the main type of DC-DC transformer. Each DAB is connected to a full-bridge converter with the AC-DC converter, and then, all the DAB converters' outputs are paralleled to feed the load.

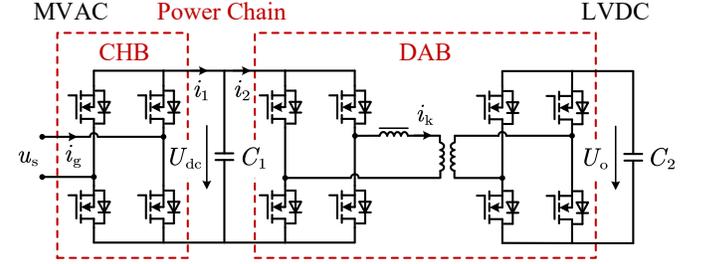

**Fig. 2.** The schematic of the solid-state transformer.

### C. DC Computing and Cooling Load

The computing load of a data center refers to the active power demand from IT equipment—including servers, storage devices, and network switches—during the execution of computational tasks [29]. It exhibits strong variability and stochastic behavior due to the dynamic nature of IT workloads. It is highly correlated with service requests, processor utilization, and virtual machine scheduling mechanisms. In general, the IT load fluctuates independently of environmental factors and tends to show a combination of periodicity and burst characteristics. Its dynamic behavior can be expressed as:

$$P_{IT}(t) = P_{base} + \alpha \cdot U(t) \quad (1)$$

where $P_{base}$ is the standby power (nearly 40-50% of rated), which shows a high baseline characteristic; $U(t)$ and $\alpha$ are the CPU/GPU utilization ratio and linear coefficient related to workload intensity, respectively.

Correspondingly, cooling load refers to the thermal energy removal required to maintain allowable temperature and humidity conditions for IT equipment operation [30], which is strongly coupled with the IT load but also influenced by environmental conditions such as outdoor temperature and humidity. Its response has thermal inertia, making cooling load changes relatively slower and smoother. The cooling load can be modeled by:

$$\tau \frac{dQ_{cool}(t)}{dt} + Q_{cool}(t) = k_1 \cdot P_{IT}(t) \quad (2)$$

where $k_1$ and $\tau$ are the heat-to-cooling conversion coefficient related to PUE and the thermal time constant, respectively.

Therefore, in our simulation framework, the cooling demand is fully incorporated and modeled in parallel with the DC computing load, together forming the total load imposed on the system.

### D. Energy Storage Systems and Distributed Photovoltaic Sources

With the increasing energy demand and environmental impact of large data centers, integrating on-site renewable generation and storage has emerged as a promising strategy to improve sustainability, reduce operational costs, and enhance grid resiliency. In particular, combining distributed photovoltaic (PV) systems with energy storage systems (ESS) enables data centers to smooth power fluctuations, reduce peak demand charges, and support continuous operation even under unstable grid or renewable conditions [31].

With the objective of directly feeding DC loads and reducing dependence on the AC grid, the PV and ESS systems are both connected in parallel to the load in the architecture, which is shown in the Fig. 3. Coordinated control mechanisms are applied to maintain reliable energy input to the load during dynamic transients.

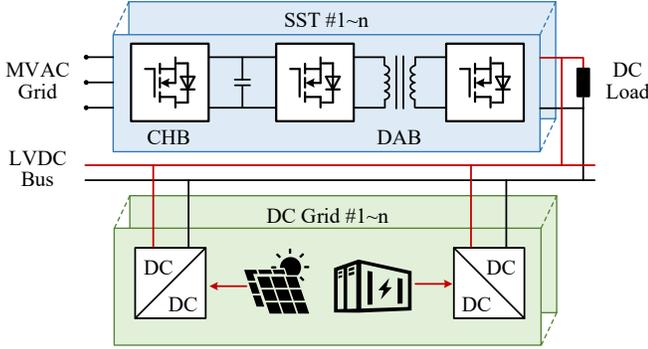

**Fig. 3.** The schematic of ESSs and PV.

## III. MODELING IMPLEMENTATION IN RTDS

Based on the overall system topology presented in Section II, we extract a single line as a representative case to develop a simulation model of a SST-driven 800 VDC data center using the high-fidelity Real-Time Digital Simulator. The model simulates the complete power transfer chain from a 10 kV AC utility supply to the 800 VDC loads. This section primarily focuses on analyzing the operational principle and control strategies of the system's key components as follows.

### A. Modeling of AC-DC Converter

The AC-DC converter is a modular multilevel converter connected to a three-phase AC grid and a DAB converter, using the component "rtds_ss_UCM_LEV2" and setting the parameter "src_ipt", which means the input source of the converter as "Modulation Waveform" in order to accept the modulation wave signal. The control strategy of the converter can be divided into High-Level Control and Low-Level Control:

*1) High-Level Control*

The high-level control for the AC-DC converter utilizes the control variables in per-unitization, which provides more convenience to modify system specifications without much change in control parameters. To start the implementation of closed control loops, the preparation of control variables and a phase-locked loop (PLL) is required. Fig. 4 shows the PLL control and DQ decompositions of the three-phase voltages and currents, which are prepared for the control of the capacitor voltages.

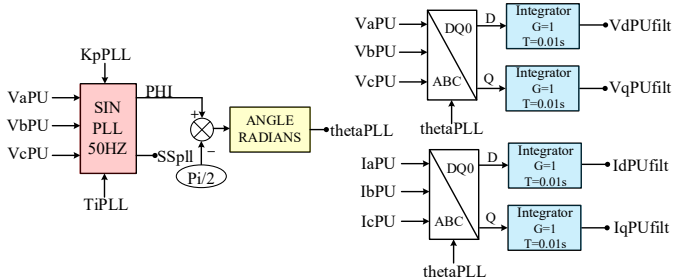

**Fig. 4.** PLL control and DQ decompositions.

After that, these control variables are sent to the closed-loop control circle to regulate the full-bridge capacitor voltages in the AC-DC converter. The closed-loop control framework hierarchically consists of an outer loop and an inner loop. The outer loop compares the measured capacitor voltage with its corresponding reference value and calculates the error, which is then fed into a PI controller to generate the reference current as the output, as seen from Fig. 5(a), at the same time, by converting the DQ-axis current reference values output by the PI controller, the corresponding amplitude and phase are derived. The amplitude is then limited to prevent it from exceeding the maximum and minimum values preset by the system, and finally, the converted DQ-axis current reference values are output. What is more, we set the IqRef manually to zero to avoid injecting unnecessary reactive power into the DC-DC converter. Fig. 5(b) shows the inner-loop decoupled current control to obtain the modulation waveform.

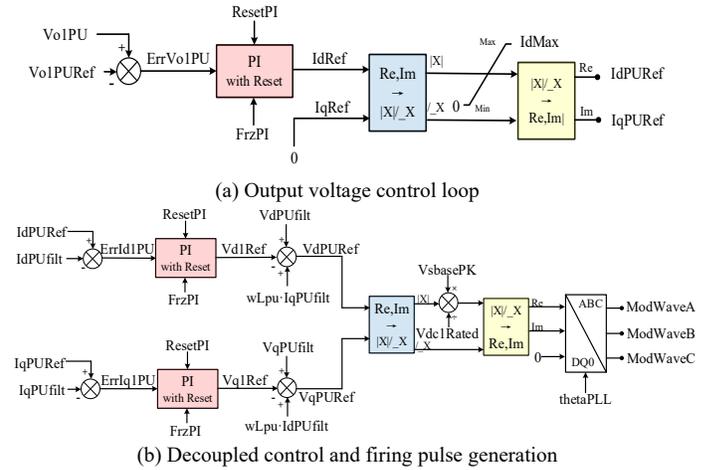

**Fig. 5.** Inner and outer loop control of AC-DC converter.

*2) Low-Level Control*

In contrast, Low-Level Control directly governs the switching of power electronic devices, such as IGBTs or MOSFETs, to accurately track the reference signals from High-Level Control. In our simulation, it is used to balance the capacitor voltages in the full bridges connected in a string, and firing pulse generation for all full-bridges. Each full-bridge capacitor voltage has its own PI control loop to generate incremental variables. This incremental variable is added to the modulation waveform obtained from high-level control.

### B. Modeling of DAB Converter

The DAB converter consists of two H-bridges on two sides of the transformer. For each H-bridge converter, it can operate with two- (i.e., +Vin and -Vin) or three-level output (i.e., +Vin, 0, and -Vin), where Vin is the input voltage of the DAB converter. For two-level output, the square-wave operation is commonly used due to the elimination of DC offset in the voltage across the transformer's windings. For three-level output, phase-shift control for switches in the H-bridge is required, which is more complicated than the control for two-level square-wave output. In this experiment, emphasis is placed on the overall system stability, control ability, and the feasibility of simulation. Employing a 2-level DAB converter markedly simplifies the system model, reduces control complexity, minimizes simulation inaccuracies, and facilitates both implementation and debugging within the RTDS envi-ronment. Therefore, the simple square-wave operation for the DAB converter is considered.

*1) Active power transmission in the DAB converter*

Simplifying the two H-bridges in the DAB converter as square-wave voltage sources, the equivalent circuit of the DAB converter is displayed in Fig. 6, where the transformer is eliminated by reflecting the secondary side to the primary side, and $L_{DAB}$ is the equivalent leakage inductance of the transformer. According to the equivalent circuit, the real power of the DAB converter delivered from the left side to the right side[32] is calculated as:

$$P_{DAB} = \frac{V_{in} \cdot V_{out} \cdot n_{tr}}{2\pi \cdot f_{sw} \cdot L_{DAB}} \cdot \varphi \cdot \left(1 - \frac{|\varphi|}{\pi}\right) \quad (3)$$

where $V_{in}$ and $V_{out}$ are the dc voltages at the left and right sides of the DAB converter, respectively; $f_{sw}$ is the switching frequency of the DAB converter; $n_{tr}$ is the turns ratio of the transformer and equals $V_{tr1}/V_{tr2}$; $\varphi$ is the phase-shift angle in radians between the primary-side and secondary-side square-wave voltages of the transformer. As indicated by (3), the power is delivered from the left to the right sides when $\varphi > 0$, and vice versa. In summary, the power is always delivered from the leading-phase side to the lagging-phase side.

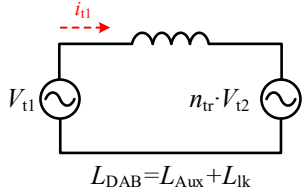

**Fig. 6.** Equivalent circuit of DAB converter.

*2) Selection of Transformer's Parameters*

As noticed from (3), the power delivered by the DAB converter is related to the transformer's parameters. To properly operate the DAB converter, the constraints or considerations are listed as follows:

i) DAB's real power must be less than the transformer's rated capacity to avoid overload;

ii) Considering all the inductance in the DAB converter is provided by the transformer's leakage inductance.

According to the above conditions, the following equations are obtained:

$$P_{DAB} = \frac{V_{in} \cdot V_{out} \cdot n_{tr}}{2\pi \cdot f_{sw} \cdot L_{lk}} \cdot \varphi \cdot \left(1 - \frac{|\varphi|}{\pi}\right) < S_{tr} \quad (4)$$

$$V_{in} = k_1 \cdot V_{tr1} \quad V_{out} = k_2 \cdot V_{tr2} \quad (5)$$

$$X_{lk} = 2\pi \cdot f_{sw} \cdot L_{lk} = \frac{X_{tr} \cdot V_{tr1}^2}{S_{tr}} \quad (6)$$

With the set of equations from (4) to (6), the following derivations can be obtained.

$$\frac{k_1 \cdot k_2}{X_{lk}} \cdot \varphi \cdot \left(1 - \frac{|\varphi|}{\pi}\right) < 1 \rightarrow X_{lk} > k_1 \cdot k_2 \cdot \varphi \cdot \left(1 - \frac{|\varphi|}{\pi}\right) \quad (7)$$

In our simulation, $k_1 = k_2 = 1$, the leakage inductance can be:

$$X_{tr} > \varphi \cdot \left(1 - \frac{|\varphi|}{\pi}\right) \quad (8)$$

All these conditions are used in the simulation.

*3) Controls of DAB Converter*

As discussed above, to control the DAB converter's power or output DC voltage, we can adjust the phase-shift angle between the voltages generated by the H-bridges connected to the transformer.

The original control strategy of DAB in RTDS uses the component "rtds_ss_phsftpwm_fpword" to generate firing pulses. Now, with the "rtds_ss_UCM_DAB_SC" model, it can receive improved firing pulses and provide better accuracy.

The improved firing pulses can be generated by using the "Internal" option for the firing pulse source configured in the UCM(Universal Converter Model)-DAB models. With the "Internal" firing pulse option, the UCM-DAB model receives the phase shift reference signal from the controller and implements the improved firing pulses based on the base frequency configured for the transformer in the DAB model. Improved firing pulses are very important to DABs, as inaccurate firing pulses can easily excite DC offset in the transformer currents. The control loop to adjust the DAB's output DC voltage is shown in Fig. 7. Similarly, a PI controller is utilized to regulate the secondary-side output voltage. Unlike the AC-DC converter, this approach omits the DQ decoupling of voltage and current, thereby simplifying the control implementation.

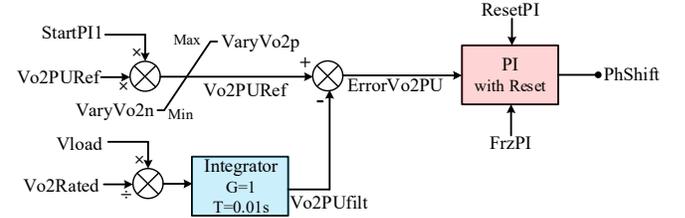

**Fig. 7.** Control loop of output DC voltage for DAB converter.

*C. Operating Controls of ESSs and Distributed PVs*

The ESS is modeled as a controllable bidirectional power unit, capable of both charging and discharging to maintain DC bus stability. Its state-of-charge (SOC) dynamics can be expressed as:

$$\frac{d}{dt}SOC(t) = -\frac{\varepsilon_{ch}P_{ch}(t) - \frac{1}{\varepsilon_{dis}}P_{dis}(t)}{E_{rated}} \quad (9)$$

where $P_{ch}$ and $P_{dis}$ are charging and discharging power; $\varepsilon_{ch}$ and $\varepsilon_{dis}$ are charging and discharging efficiencies; $E_{rated}$ is the rated energy capacity.

The distributed photovoltaic subsystem operates primarily as a controlled DC power source interfaced to the DC bus through a DC-DC converter. The fundamental control objective is to regulate the PV-side operating point to maximize power extraction while ensuring stable DC-link regulation. A commonly adopted strategy is the Maximum Power Point Tracking (MPPT) to adjust the PV array voltage. When the PV output exceeds the load demand, surplus energy is directed to the ESS for charging; otherwise, the PV contributes partially or fully to the instantaneous load.

Since this experiment primarily addresses the sequential operation of the system under steady-state conditions and does not consider transient processes, photovoltaic generation and energy storage are not reflected in the simulation.

## IV. SIMULATION RESULTS AND ANALYSIS

### A. Simulation Settings

This section describes two models of data center: a traditional model with UPS power infrastructure and a model with SST. To simulate converters with a high switching frequency, the DC-DC transformer simulations are implemented in the Substep environment in small-dt boxes. The corresponding parameter settings are shown in Table I.

TABLE I
SYSTEM PARAMETER SETTINGS

| No. | Parameter | Value |
| --- | --- | --- |
| 1 | System fundamental frequency | 50 Hz |
| 2 | Switching frequency for full-bridge converters | 200 Hz |
| 3 | Output capacitance of full-bridge converter | 2700 μF |
| 4 | AC-side inductance as filter | 0.015 H |
| 5 | Line-to-line RMS voltage of AC grid | 10.0 kV |
| 6 | Rated capacitor voltage | 1.5 kV |
| 7 | Rated output real power | 2.25 MW |
| 8 | Switching frequency | 1000 Hz |
| 9 | Transformer rated capacity per DAB | 0.75 MW |
| 10 | Primary-side rated voltage | 1.5 kV |
| 11 | Secondary-side rated voltage | 1.5 kV |
| 12 | Transformer leakage inductance (p.u.) | 0.376 pu |
| 13 | Input capacitance | 2000 μF |
| 14 | Rated output DC voltage | 0.8 kV |
| 15 | Output capacitance | 50 μF |
| 16 | Rated output resistive load | 1 Ω |

### B. Architecture Benchmarks

The UPS power supply architecture consists of a multi-stage conversion chain designed to ensure stable and uninterrupted power delivery to downstream IT equipment, which is shown in the Fig. 8.

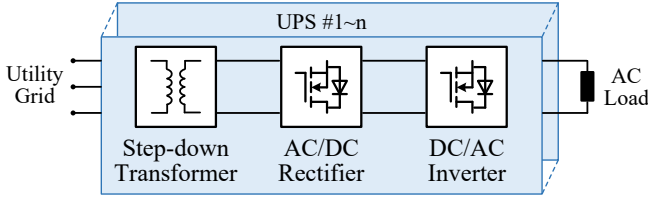

Fig. 8. Power Supply Architecture Diagram of UPS.

The process begins with an inverter stage, which converts the incoming DC source into AC power for subsequent conditioning. The AC output is then fed into a step-up transformer that elevates the voltage to a suitable medium-voltage level for efficient distribution. Following this, a step-down transformer reduces the voltage to a level appropriate for rectification. The AC/DC rectifier converts the transformed AC power back into DC, providing a stable DC link for the final stage. Lastly, a DC/AC inverter generates regulated AC power for the load, ensuring consistent voltage and frequency even under grid disturbances or upstream fluctuations. This multi-stage architecture enhances power quality and reliability, forming the core of traditional double-conversion UPS systems widely used in data-center environments.

### C. Operating Performance Analysis

To evaluate the performance of the proposed SST-based power supply architecture under realistic data-center conditions, time-domain simulations are carried out using measured load profiles and are compared with a benchmark UPS-based architecture. The two architectures share the same system parameters and control settings; only the power conversion topology and the load time series are changed.

Three types of operating scenarios are considered.

i) Typical-day AI computing load: A 24-hour power profile is taken from an AI cluster and represents the intra-day fluctuation of computing tasks. The measured power sequence is applied to the 800 VDC bus with the original sampling interval $\Delta t$.

ii) Month-long operation with realistic fluctuations: Thirty consecutive days of measured power data are used. Two kinds of load are distinguished: AI data-center load (AIDC) and conventional IT data-center load (IDC). For each day, the same power sequence is fed into the SST and the UPS architectures separately so that both systems experience identical load trajectories. The resulting monthly input energy and power-quality indices are then obtained.

iii) Low-fluctuation reference load: To examine the impact of short-term power swings, a smoothed reference profile is constructed from the AIDC data. For each day, the deviation of instantaneous power from the daily average is compressed by a scaling factor while the daily energy is kept unchanged. In this way, the long-term energy requirement is preserved but the fast power variation is significantly reduced, which mimics a more traditional IT load.

For each scenario, the following metrics are evaluated.

i) Energy-efficiency: The electrical energy drawn from the upstream grid over the simulation period is calculated and used to compare the SST and UPS architectures under the same load energy demand.

ii) DC-bus voltage: For the 800 VDC bus, the mean value $U_{avg}$, minimum $U_{min}$, maximum $U_{max}$ are recorded. The standard deviation $\sigma_U$ is calculated as:

$$\sigma_U = \sqrt{\frac{1}{N}\sum_{i=1}^{N}(u_i - U_{avg})^2} \quad (10)$$

And the relative percentage deviation $\delta_U\%$ is calculated as:

$$\delta_U\% = \frac{U_{avg} - U_{ref}}{U_{ref}} \times 100\% \quad (11)$$

where the $U_{ref}$ is the rated output DC voltage: 0.8 kV. The percentages of time during which the DC voltage stays within ± 1%, ±2% of the nominal value are also calculated as:

$$within_1 = \frac{|\{i \,||\, |u_i - U_{ref}| \le tol_1\}|}{N} \times 100\% \quad (12)$$

$$within_2 = \frac{|\{i \,||\, |u_i - U_{ref}| \le tol_2\}|}{N} \times 100\% \quad (13)$$

where $tol_1$ and $tol_2$ are the tolerance bounds, which can be calculated as:

$$tol_1 = 0.01 U_{ref} \quad tol_2 = 0.02 U_{ref} \quad (14)$$

We also consider the Ripple Factor: the Peak-to-Peak Ripple Coefficient $Kr_{pp}$ is calculated as:

$$Kr_{pp} = \frac{U_{max} - U_{min}}{U_{avg}} \quad (15)$$

And the RMS Ripple Coefficient $Kr_{rms}$ can be calculated as:

$$Kr_{rms} = \frac{u_{ac,rms}}{U_{avg}} \quad (16)$$

where $u_{ac,rms}$ comes from:

$$u_{ac,rms} = \sqrt{\frac{1}{N}\sum_{i=1}^{N}(u_{ac,i})^2} \quad (17)$$

And the $u_{ac,i}$ comes from:

$$u_{ac,i} = u_i - U_{avg} \quad (18)$$

iii) Harmonic and power-quality: The three-phase AC bus voltages are analyzed by the Fourier transform. The harmonic content and total harmonic distortion (THD) are obtained, and for month-long simulations, the mean value and standard deviation over 30 days are calculated.

The following sections analyze the predefined operating scenarios individually.

*1) Typical-day AI load: DC-bus and loss behavior*

The analysis begins with a typical-day AI computing load. The 24h measured power sequence is applied to the 800 VDC bus, and the corresponding DC-bus voltage and input power responses of the SST architecture are obtained. The AI computing load curve for this day is shown in Fig. 9.

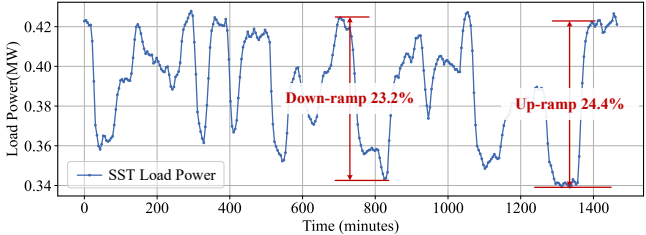

**Fig. 9.** The 24-hour measured load power of the AI cluster.

Fig. 10 presents the time series of the 800 VDC bus voltage over a typical day. Although the AI load exhibits pronounced peaks and valleys, the DC-bus voltage remains tightly regulated around the nominal value. The statistical indices in Table 2 show that the voltage variation is very small. Further statistics indicate that the voltage stays within ±2% of the rated value for 100% of the time.

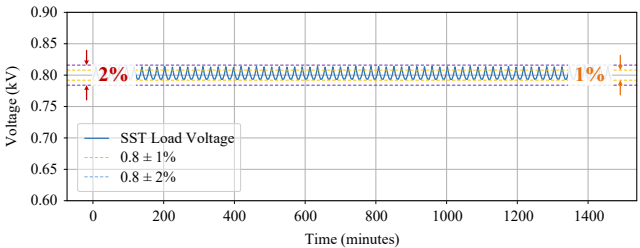

**Fig. 10.** 800 VDC bus voltage.

In other words, under the dynamic impact of the typical-day AI load, the SST-based architecture is still able to maintain the 800 VDC bus voltage within a very narrow band. Short-term fluctuations are almost completely absorbed by the DC-side control and energy-storage stages. This shows that, for the typical-day AI operating condition, the SST architecture can effectively suppress power fluctuations of the computing load and decouple them from the upstream grid.

TABLE II
THE DAILY STATISTICAL INDICES OF 800 VDC BUS VOLTAGE

| No. | Parameter | Value |
|---|---|---|
| 1 | $U_{avg}$ | 0.80040 |
| 2 | $U_{min}$ | 0.79285 |
| 3 | $U_{max}$ | 0.81479 |
| 4 | $\sigma_U$ | 0.00726 |
| 5 | $\delta_U\%$ | 0.05010% |
| 6 | $Kr_{pp}$ | 0.02741 |
| 7 | $Kr_{rms}$ | 0.00907 |
| 8 | $within_1$ | 80% |
| 9 | $within_2$ | 100% |

Fig. 11 shows the instantaneous power-loss curve $P_{loss}(t)$ for the same typical day, the difference between input power and load power as a function of time. The loss remains at a relatively low level throughout the day and only increases briefly when the load power rises or falls rapidly. During other periods, the loss fluctuates mildly and is concentrated within a narrow range. Overall, the variation of the loss is not synchronized with the aggressive power swings of the AI load, which means that the DC-side energy buffering and peak-shaving functions do not introduce significant additional loss.

Together with the subsequent daily and monthly statistics, this indicates that the system can maintain a stable DC-bus voltage while still achieving relatively high energy efficiency. The corresponding daily power-loss curve illustrates this behavior of the instantaneous loss more clearly. Compared with Fig. 9, we can find that the load power curve and the loss curve show opposite trends. This behavior may result from the current–voltage matching strategy adopted in the simulation model to optimize voltage control. When the load increases, the PI controller adjusts the current to maintain stable voltage. Since the loss is proportional to the square of the current, the controller may cause the decrease in current to be larger than the increase in load, resulting in the loss curve showing a trend that changes in the opposite direction of the load.

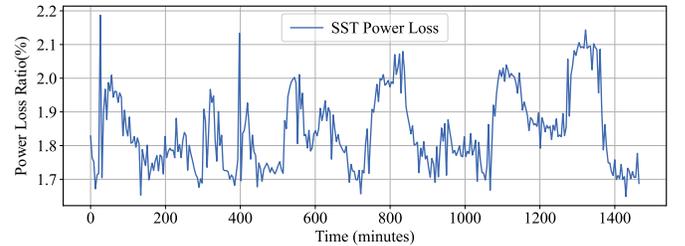

**Fig. 11.** The instantaneous power-loss curve in a typical day.

*2) Month-long operation with AI loads: SST versus UPS*

Based on the typical-day analysis, a month-scale simulation is carried out using 30 consecutive days of measured data. For each daily power sequence, the SST and UPS architectures are simulated separately so that both systems are subjected to exactly the same load fluctuations, as shown in Fig. 12.

Under the 30-day realistic operating condition, Fig. 13(a) shows the input-power time series of the SST-based and UPS-

based supply architectures. The blue curve corresponds to the SST and the orange curve to the UPS. Both curves exhibit obvious daily peak-valley patterns following the load, but over the entire simulation period, the UPS input power is consistently higher than that of the SST. The UPS power rises more sharply during peak periods, and in low-load periods, it still remains on a higher plateau. In contrast, the SST power fluctuates within a lower range with relatively moderate amplitude. This indicates that, under the same AI load-energy demand, the SST architecture generally draws less power from the grid throughout the operating period.

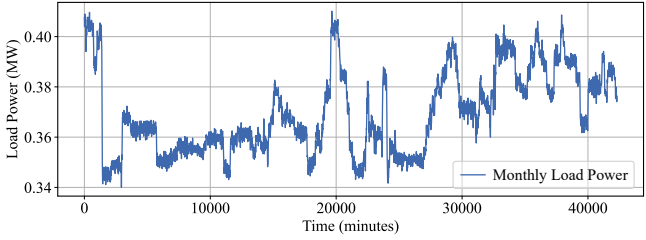

**Fig. 12.** Monthly Load Curve.

Fig. 13(b) shows the corresponding instantaneous power-loss curve $P_{\text{loss}}(t)$, again with blue representing the SST and orange representing the UPS. Similar to the input power, the loss curve of the UPS remains at a higher level and exhibits strong variation with the load, whereas the loss of the SST is significantly lower and confined to a narrower band. In other words, when the load fluctuates strongly, the UPS not only requires higher input power but also generates larger actnal conversion losses, while the SST, through more efficient power conversion and DC-side energy buffering, keeps the losses at a relatively low level.

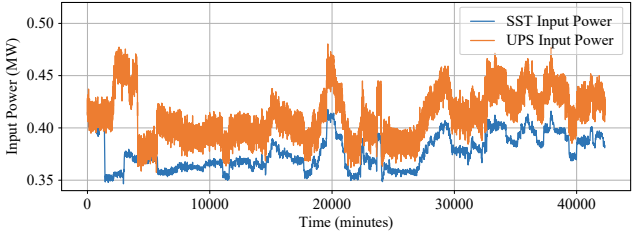

(a) Grid-side power curves

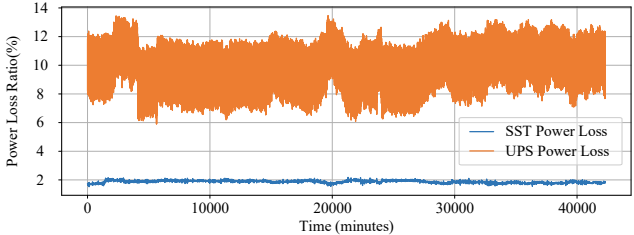

(b) Power loss curves

**Fig. 13.** Power Amplitude Diagram for Different Schemes.

Table III lists the total input average loss of the two architectures over the 30-day simulation period: 1.924% for the SST-based architecture and 9.553% for the UPS-based architecture. The difference is about 7.629%, which means that the SST draws roughly 8.5% less energy from the grid than the UPS. Combined with the observation from Fig. 13(a) that the SST has a lower input-power level, and from Fig. 13(b) that its instantaneous loss is considerably smaller, it can be concluded that, under typical month-long realistic operating conditions, the SST-based architecture can significantly reduce system losses and input energy while delivering the same computing service. It therefore exhibits better energy efficiency and operating economy than the conventional UPS-based architecture.

TABLE III
MONTHLY INPUT-SIDE AVERAGE LOSS

| No. | Power Supply Architecture | Average Loss Ratio |
|---|---|---|
| 1 | SST | 1.924% |
| 2 | UPS | 9.553% |

After obtaining the active power curve of the input, we perform a frequency domain analysis. As both the power grid and the units themselves have various "natural frequencies," if the power frequency of the load coincides with these natural frequencies, it may trigger resonance in the upstream units. Over time, this can lead to shaft fatigue, even shaft breakage, or cause large-scale oscillations and separation of power grids. Thus, utility companies typically provide data centers with a frequency-domain specification, specifying a "critical frequency range," e.g., 0.1-20 Hz. Within this range, some metrics of the power spectrum (e.g., normalized magnitude) must not exceed certain limits. In other words, the "power spectrum" of the GPU must be sufficiently low at these frequencies to avoid exciting the resonance modes of the power grid.

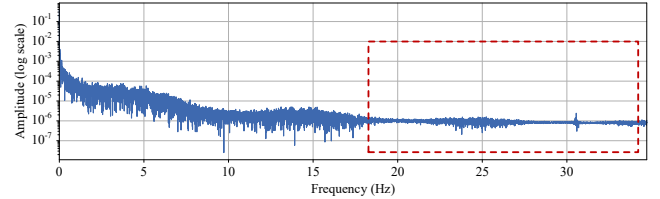

(a) SST

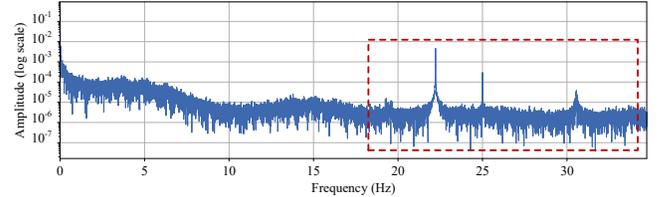

(b) UPS

**Fig. 14.** Frequency components of the power waveform.

The source-side power spectra of the two AIDC supply schemes show that the UPS-based architecture exhibits several sharp low-frequency peaks, indicating strong periodic power swings concentrated at a few dominant frequencies. This narrowband behavior is prone to interacting with grid electromechanical modes. In contrast, the SST-based architecture significantly lowers the spectral magnitude across the 0.1–a few hertz range and yields a more broadband, flattened spectrum. This reflects smoother power injection at the point of common coupling and a reduced tendency to excite critical low-frequency resonances, demonstrating the superior grid-friendliness of the SST scheme.

For the 30-day realistic fluctuation scenario, the statistical indices of the 800 VDC bus voltage are consistent with those of the typical day. Compared with the typical-day case, the monthly results show almost the same average value and only slightly larger standard deviation and ripple factors, with increases within 10%, all at a very low level.

TABLE IV
THE MONTHLY STATISTICAL INDICES OF 800 VDC VOLTAGE

| No. | Parameter | Value |
|---|---|---|
| 1 | $U_{avg}$ | 0.80042 |
| 2 | $U_{min}$ | 0.79264 |
| 3 | $U_{max}$ | 0.81480 |
| 4 | $\sigma_U$ | 0.00738 |
| 5 | $\delta_U\%$ | 0.04525% |
| 6 | $Kr_{pp}$ | 0.05337 |
| 7 | $Kr_{rms}$ | 0.02766 |
| 8 | $within_1$ | 80% |
| 9 | $within_2$ | 100% |

Comparing the two sets of results leads to two direct conclusions. First, the voltage statistics of the typical-day load are highly consistent with those of the full month, which means that the selected typical day is representative and not a special case with exceptionally small fluctuations. Second, when the simulation time is extended from one day to thirty days, no evident degradation or long-term drift is observed in the mean value, standard deviation or ripple indices of the DC-bus voltage. This indicates that the voltage control of the SST-based architecture remains stable and reliable during long-term operation and does not suffer from an accumulation of control error over time.

### D. Exploratory Experiments and Insights

#### 1) Exploration of load fluctuation effects

To further clarify how load fluctuation characteristics influence the supply architectures, a low-fluctuation reference load is constructed from the month-long AI load. Specifically, using the monthly AI load as the basis, the instantaneous power deviation from the daily average is scaled down for each day, while a correction factor is used to keep the daily energy unchanged. The resulting reference load has the same long-term energy requirement as the original AI load but significantly reduced short-term fluctuations, making it more similar to a traditional IT load. We can see the difference in Fig. 15.

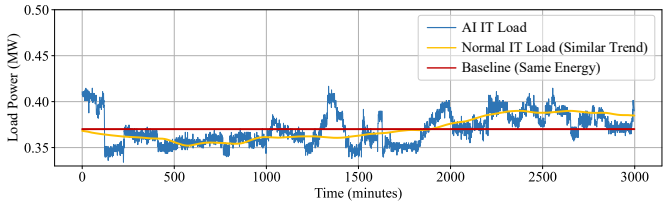

**Fig. 15.** Traditional IT Load Power Curves.

To compare AI loads and conventional IT loads, Fig. 16 extends the monthly input-energy statistics to four operating cases: AI-UPS, AI-SST, conventional-UPS and conventional-SST. From the bar chart it can be seen that, for the same AI load, the SST solution consumes about 26.36 MWh less energy than the UPS, corresponding to an energy-saving rate of about 9.0%; for the conventional load, the SST solution consumes about 20.90 MWh less energy than the UPS, corresponding to an energy-saving rate of about 7.27%. For a given supply architecture, the monthly input energy under the AI load is slightly higher than that under the conventional load. This difference is more significant in the UPS architecture, while in the SST architecture, the two are closer, which suggests that the conventional UPS is more sensitive to AI load fluctuations.

Therefore, regardless of load type, the SST architecture provides a stable energy-saving advantage of about 7–9% over the conventional UPS architecture on a monthly time scale. At the same time, the strong fluctuation of the AI load causes a more significant increase in energy consumption for the conventional UPS, whereas the SST is less sensitive to the load type and is better suited to coping with the highly fluctuating power demand of AI workloads.

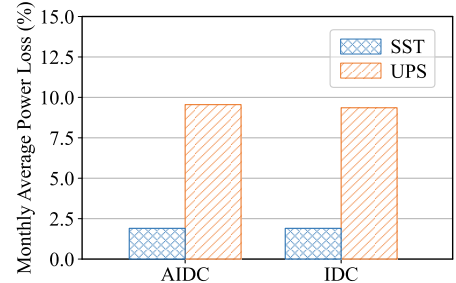

**Fig. 16.** Monthly total electricity consumption under four operating conditions.

As shown in Table V, it can be concluded that, under different load types (highly fluctuating AI loads and relatively smooth conventional IT loads), the 800 VDC bus voltage of the SST architecture exhibits almost the same statistical characteristics. The load type mainly affects the power-loss distribution and energy consumption of the conventional UPS, but does not significantly worsen the DC-bus voltage quality under the SST architecture. For the same AI load, the conventional UPS shows higher energy consumption and greater sensitivity, whereas the SST architecture maintains both lower energy use and more stable voltage quality.

TABLE V
SST SOLUTION DC VOLTAGE REGULATION CHARACTERISTICS UNDER DIFFERENT LOAD CONDITIONS

| Parameter | AIDC | IDC |
|---|---|---|
| $U_{avg}$ | 0.80042 | 0.79997 |
| $U_{min}$ | 0.79264 | 0.79160 |
| $U_{max}$ | 0.81480 | 0.81166 |
| $\sigma_U$ | 0.00738 | 0.00650 |
| $\delta_U\%$ | 0.04525% | -0.00422% |
| $Kr_{pp}$ | 0.05337 | 0.0251 |
| $Kr_{rms}$ | 0.02766 | 0.00810 |
| $within_1$ | 80% | 79.99% |
| $within_2$ | 100% | 100.0% |

As shown in Fig. 17, the SST and UPS schemes exhibit almost identical temporal patterns of input power over the entire month, clearly following the same IT load profile with pronounced daytime peaks and nighttime valleys. However, the UPS scheme consistently requires a higher input power than the SST scheme to supply the same DC load. This indicates additional conversion and standby losses associated with the multi-stage AC–DC–AC structure and energy storage interfaces in the UPS configuration, whereas the SST provides a shorter conversion chain and thus a higher efficiency.

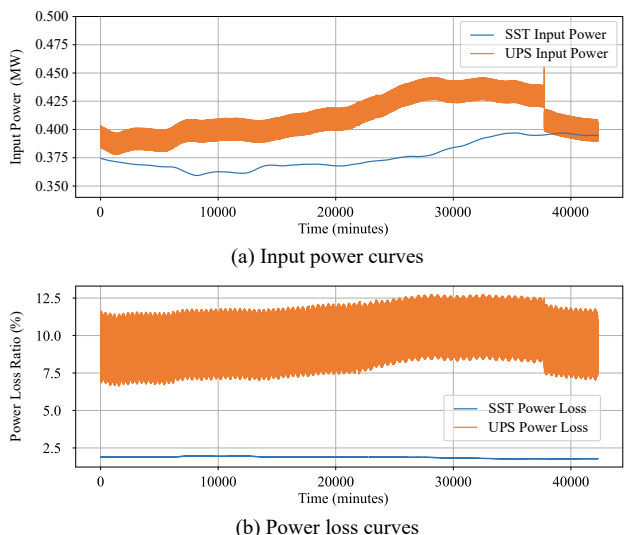

(a) Input power curves

(b) Power loss curves

**Fig. 17.** Power Amplitude Diagram for Different Schemes.

The loss curves shown in the Fig. 18 further highlight this difference in performance. For the SST scheme, the power loss ratio remains tightly clustered around approximately 2%, with only minor fluctuations as the load varies, demonstrating both high and stable conversion efficiency. In contrast, the UPS scheme operates with a substantially higher loss ratio, typically within the range of about 7%–13%, and exhibits stronger sensitivity to load level, with loss peaks occurring during high-demand periods. Superimposed high-frequency ripples in the UPS loss curve suggest additional dynamic losses related to switching and filter components. Overall, the monthly statistics confirm that the SST scheme achieves markedly lower energy losses and more stable efficiency than the conventional UPS scheme.

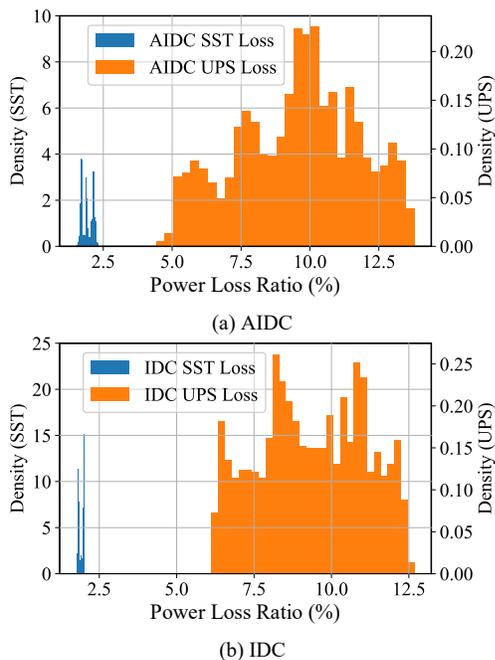

(a) AIDC

(b) IDC

**Fig. 18.** Monthly Power Error Distribution Histogram for Different Schemes.

Fig. 19 is based on statistics of the 30-day simulation, presenting the "monthly loss" in the AI-load and conventional-load cases using histograms. These figures show that, under the AI load, the daily input-energy distribution range is clearly wider than under the conventional load, indicating that the AI load leads not only to a higher overall energy level but also to larger variation and a wider energy range.

Comparing the SST and UPS for the same load type, the energy distribution of the SST is shifted downward as a whole and slightly less dispersed, which implies that in both AI and conventional scenarios, the SST architecture not only has a lower average energy consumption but also exhibits better controllability of daily fluctuations.

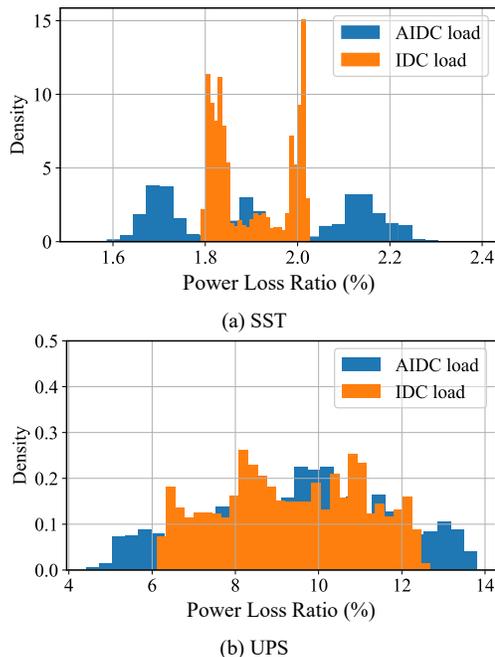

(a) SST

(b) UPS

**Fig. 19.** Monthly Power Error Distribution Histogram for Different Loads.

Overall, the simulations suggest that the SST-based power supply architecture provides stable DC-bus voltage and significant energy savings across different load types and time scales, and that its advantage becomes more evident when serving highly variable AI data-center workloads.

*2) Low-Voltage DC-Side Capacitance Analysis*

This section presents an exploratory study on the selection of the low-voltage-side capacitor in the DAB of the SST. Simulations are performed with capacitance values ranging from 30 μF to 60 μF to evaluate their effects on key power performance indices, with all other parameters held constant.

According to the results in these three figures, the influence of the DC-side capacitance (C) on the SST performance can be summarized as follows.

In Fig. 20(a), the load-side voltage ripple decreases sharply when the capacitance increases from 30 to about 55 μF, and then tends to saturate, indicating that most of the stabilizing effect is achieved within this range. Fig. 20(b) shows that the average power loss remains nearly constant over 30–60 μF, so enlarging the capacitor does not introduce a noticeable efficiency penalty. In Fig. 20(c), the spectral energy of input-power oscillations in the 0.1–3 Hz band reaches a broad minimum for capacitances around 50–60 μF, implying effective damping of low-frequency power fluctuations.

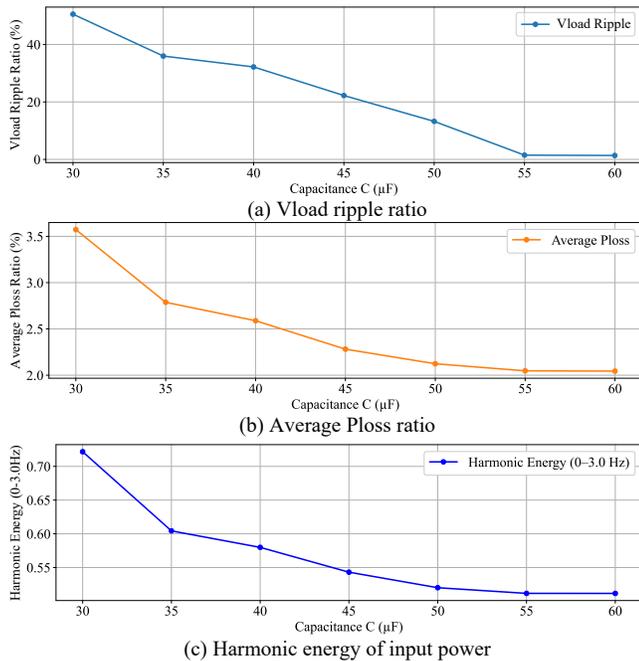

**Fig. 20.** Impact of capacitor selection on data center power supply reliability.

Combining these three indicators, 55 μF can be regarded as an appropriate compromise for the DC-side capacitance. At C=55 μF, both the load-side voltage ripple and the low-frequency power spectral energy are close to their minimum values, while further increasing brings no substantial benefit but would increase the capacitor volume, cost and inrush current. Therefore, a capacitance of approximately 55 μF is selected as the optimal design point for the SST in this study.

## V. CONCLUSION

This paper proposes an SST-driven 800 VDC power architecture for data centers, modeling a three-stage chain with a three-phase H-bridge AC/DC front end and a DAB DC/DC stage and designing coordinated closed-loop control. RTDS-based and day/month time-series simulations verify stable DC-bus regulation and acceptable power-quality performance under realistic AI load variability. Benchmarking against a conventional UPS supply chain shows a clear reduction in input-side losses for the same delivered load energy. A sensitivity study further clarifies how DC-link capacitance and load fluctuation characteristics affect voltage ripple and low-frequency input-power oscillations, providing practical capacitance sizing guidance. Overall, the work offers a reproducible evaluation workflow and actionable design insights for next-generation 800 VDC AI data centers. Future work may investigate transient protection mechanisms for 800 VDC data centers. Meanwhile, data center power supply architectures beyond the traditional 2N redundancy scheme can be explored.